\documentclass[twocolumn,preprintnumbers,superscriptaddress]{revtex4-1}

\usepackage{amsmath}
\usepackage{graphicx}
\usepackage{mathrsfs}
\usepackage{amssymb}
\usepackage{amsfonts}
\usepackage{amsmath}
\usepackage{CJK}
\usepackage{epsfig}
\usepackage{epstopdf}
\usepackage[colorlinks=true,linkcolor=blue,allcolors=blue,citecolor=blue]{hyperref}
\usepackage{dcolumn}
\usepackage{epsfig}
\usepackage{epstopdf}
\usepackage{graphicx}
\usepackage{ulem}
\usepackage[utf8]{inputenc}
\usepackage{amsthm}
\usepackage{bm}
\usepackage{color}
\usepackage{float}
\usepackage{appendix}
\usepackage{setspace}
\usepackage{geometry}
\begin{document}
\setlength{\textfloatsep}{2pt}
\setlength{\intextsep}{2pt}

\title{Phonon-laser ultrasensitive force sensor}
\author{Zhichao Liu}
\affiliation{State Key Laboratory of Magnetic Resonance and Atomic and Molecular Physics, Wuhan Institute of Physics and Mathematics, Innovation Academy of Precision Measurement Science and Technology, Chinese Academy of Sciences, Wuhan 430071, China}
\affiliation{University of the Chinese Academy of Sciences, Beijing 100049, China}
\author{Yaqi Wei}
\affiliation{State Key Laboratory of Magnetic Resonance and Atomic and Molecular Physics, Wuhan Institute of Physics and Mathematics, Innovation Academy of Precision Measurement Science and Technology, Chinese Academy of Sciences, Wuhan 430071, China}
\affiliation{University of the Chinese Academy of Sciences, Beijing 100049, China}
\author{Liang Chen}
\email{Corresponding author: liangchen@wimp.ac.cn}
\affiliation{State Key Laboratory of Magnetic Resonance and Atomic and Molecular Physics, Wuhan Institute of Physics and Mathematics, Innovation Academy of Precision Measurement Science and Technology, Chinese Academy of Sciences, Wuhan 430071, China}
\affiliation{Research Center for Quantum Precision Measurement, Guangzhou Institute of Industry Technology, Guangzhou, 511458, China }
\author{Ji Li}
\affiliation{State Key Laboratory of Magnetic Resonance and Atomic and Molecular Physics, Wuhan Institute of Physics and Mathematics, Innovation Academy of Precision Measurement Science and Technology, Chinese Academy of Sciences, Wuhan 430071, China}
\affiliation{University of the Chinese Academy of Sciences, Beijing 100049, China}
\author{Shuangqing Dai}
\affiliation{State Key Laboratory of Magnetic Resonance and Atomic and Molecular Physics, Wuhan Institute of Physics and Mathematics, Innovation Academy of Precision Measurement Science and Technology, Chinese Academy of Sciences, Wuhan 430071, China}
\affiliation{University of the Chinese Academy of Sciences, Beijing 100049, China}
\author{Fei Zhou}
\affiliation{State Key Laboratory of Magnetic Resonance and Atomic and Molecular Physics, Wuhan Institute of Physics and Mathematics, Innovation Academy of Precision Measurement Science and Technology, Chinese Academy of Sciences, Wuhan 430071, China}
\affiliation{Research Center for Quantum Precision Measurement, Guangzhou Institute of Industry Technology, Guangzhou, 511458, China }
\author{Mang Feng}
\email{Corresponding author: mangfeng@wipm.ac.cn}
\affiliation{State Key Laboratory of Magnetic Resonance and Atomic and Molecular Physics, Wuhan Institute of Physics and Mathematics, Innovation Academy of Precision Measurement Science and Technology, Chinese Academy of Sciences, Wuhan 430071, China}
\affiliation{University of the Chinese Academy of Sciences, Beijing 100049, China}
\affiliation{Research Center for Quantum Precision Measurement, Guangzhou Institute of Industry Technology, Guangzhou, 511458, China }
\affiliation{School of Physics, Zhengzhou University, Zhengzhou 450001, China}

\begin{abstract}
Developing nano-mechanical oscillators for ultrasensitive force detection is of great importance in exploring science. We report our achievement of ultrasensitive detection of the external force regarding the radio-frequency electric field by a nano-sensor made of a single trapped $^{40}$Ca$^{+}$ ion under injection-locking, where squeezing is additionally applied to detection of the smallest force in the ion trap. The employed ion is confined stably in a surface electrode trap and works as a phonon laser that is very sensitive to the external disturbance. The injection-locking drove the ion's oscillation with phase synchronization, yielding the force detection with sensitivity of 347 $\pm$ 50 yN/$\sqrt{Hz}$. Further with 3 dB squeezing applied on the oscillation phase variance, we achieved a successful detection of the smallest force to be 86.5 $\pm$ 70.1 yN.
\end{abstract}
\maketitle
%\begin{spacing}{1.0}
\section{INTRODUCTION\vspace{-1em}}
With development of nanotechnology using atomic and molecular sensors, the physical quantities have been detected in a higher level of sensitivity than ever before due to the excellent perceived performance of the sensors \cite{Hosten2016,Schreppler2019,Mamin2001,Lecocq2015,Caves1980,Degen2017,Kotler12011,Kim2017}. Force detection has been always taking an important role in precision measurement since all the changes of motion are concerned with the force due to Newton's laws, where the small force detection by trapped ions or atoms has recently demonstrated the superiority of high sensitivity and broad adjustability \cite{Gilmore2017,Affolter2020,Gilmore2021,Shaniv2017,Biercuk2010,Blums2018,Araneda2019,Knunz2010}. The weak AC electric force was sensitively detected by means of spin-motion entanglement in two-dimensional trapped-ion mechanical oscillators \cite{Gilmore2017,Affolter2020,Gilmore2021}. The radio-frequency electric force as small as 5 yN was measured, based on the technique of injection-locked phonon laser, by observing the range of injected frequency and the gravity variance of the ion's displacement \cite{Knunz2010}. Such a radio-frequency electric force was also detected by phase-coherent Doppler velocimetry, reaching the sensitivity of 390 $\pm$ 150 yN/$\sqrt{Hz}$ \cite{Biercuk2010}. A recent detection of the sub-attonewton force regarding DC electric field was accomplished by measuring the ion's displacement three-dimensionally at nanometer precision \cite{Blums2018}.

In the present work, we report our ultrasensitive detection of the radio-frequency electric force at hundred yoctonewton's order of magnitude by a nano-probe made of a single trapped $^{40}$Ca$^{+}$ ion. Although our detection, similar to in \cite{Knunz2010}, also takes advantage of the stably confined ion behaving as a phonon laser, we acquired the sensitivity of the sensor and the lower bound of the small force by faster measurements compared to the method of scanning the injection frequency in \cite{Knunz2010}. In addition, squeezing technique \cite{Natarajan1995} was employed to further suppress thermal noises in the system.
The ion's phonon laser under our consideration works as an amplitude-amplified harmonic oscillator, which is quite sensitive to the external force, as reported previously \cite{Vahala2009,Grudinin2010,Ip2018,He2016,Hush2015,Dominguez2017}. In our experiment, the trapped $^{40}$Ca$^{+}$ ion was first Doppler-cooled and then pumped simultaneously by a red-detuned laser and a blue-detuned laser with respect to the resonant transition. After locking the oscillation frequency by injection signal, we might detect the injected force, which is induced by a radio-frequency electric field, based on the photon count measurements of the oscillation amplitude, the latter of which was carried out by fitting the fluorescence curve directly acquired from the photomultiplier tube (PMT) along with the synchronized measurement of the injection-locking signal. Our achieved force sensitivity is as low as 347 $\pm$ 50 yN/$\sqrt{Hz}$, and the oscillation amplitude uncertainty is only 15 nm with respect to the oscillation amplitude $\approx$18 $\mu$m. The detection sensitivity represents the minimum detectable change rate of the force. In contrast, we may also try to detect the smallest force, indicating the minimum force our probe could sense. To this end, we introduce the squeezing technique as in \cite{Natarajan1995} and with 3 dB squeezing applied on the oscillation phase variance, we detected successfully the smallest force of 86.5 $\pm$ 70.1 yN. Our technique could be applied to the oscillation force sensors using other ions, atoms or solid-state oscillators.

\begin{figure*}
\includegraphics[width=16 cm ]{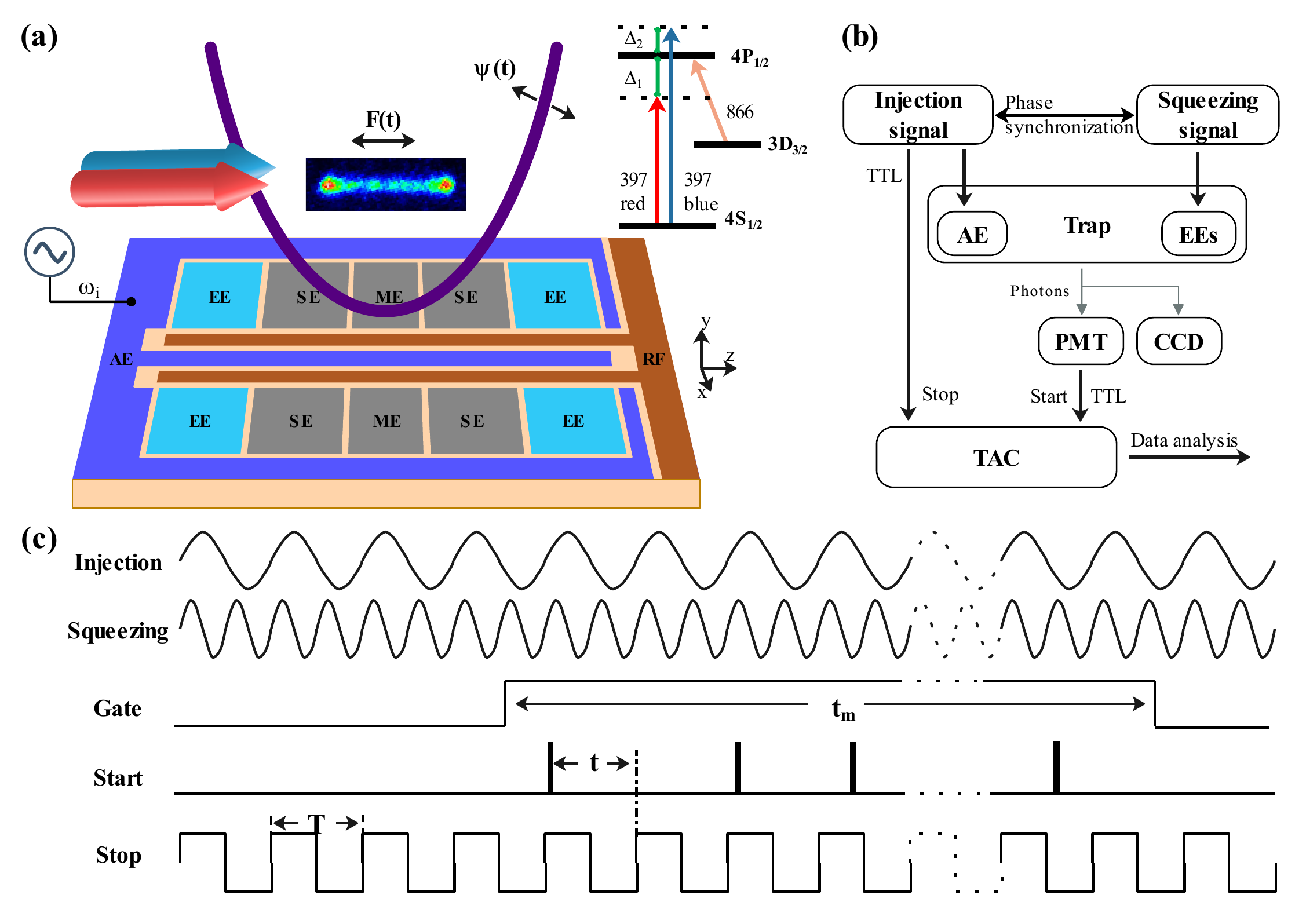}
\caption{Experimental system and scheme. (a) Sketch of the phonon laser of an ion trapped above the surface of the SET. The $^{40}$Ca$^{+}$ ion with oscillating motion irradiated by red- and blue-detuned 397-nm laser beams is trapped by the potential $\Psi(t)$ regarding a time-varying electric field. The electrodes are labelled with names for convenience of description in the text, where EE, SE, ME, AE and RF indicate, respectively, the endcap electrode, the steering electrode, the middle electrode, the axial electrode and the radio-frequency electrode. The injection signal applied on AE with frequency $\omega_{i}=\omega_{z}$, locking the oscillation frequency, is to be detected. The squeezing, with frequency doubling of $\omega_{i}$, acts on EEs when required. The top-right inset is the level scheme of $^{40}$Ca$^{+}$ for our purpose. (b) Signal processing scheme of the photon-arrival-time measurements. The applied injection signal and the squeezing signal are phase synchronized, with the latter of frequency doubling with respect to the former. The photon-arrival-time transited to voltage is measured by TAC. The ratio of the beam splitter for the PMT to CCD is set to be 7/3. (c) Signal and sequence diagram. The Injection signal contributes to locking the oscillation frequency/phase. The Squeezing signal is applied, when required, to suppress the variance of the oscillation phase. The Gate signal controls the measurement time $t_{m}$ containing a series of oscillation period $T$ and the TAC records the photon's arrival time $t$ in each period. The Start signal comes from the PMT regarding the photon pulses. The Stop TTL is synchronized with injection locking signal and the Stop signal is produced by the photon pulses from the spontaneous emission. For clarity, the plotted pulses are not strictly to scale. The typical values: $t_{m}$ = 10 s, $T$ = 5.38 $\mu$s, and 0 $<~t~\leqslant~T$. }
\label{fig1}
\end{figure*}

\section{Experimental system and scheme\vspace{-1em}}
Our experiment of ultrasensitive force detection is carried out by an injection-locked phonon laser regarding a single trapped ion confined stably in the harmonic potential of a surface-electrode trap (SET), see Fig. \ref{fig1}a. The SET employed in our experiment was introduced previously \cite{Brownnutt2015,Wan2013,Liu2020,House2008}, which is a 500-$\mu$m scale planar trap composed of two pairs of endcap electrodes, two pairs of steering electrodes, a pair of middle electrodes, an axial electrode and a radio-frequency electrode. The measured secular frequencies of the SET are, respectively, $\omega_{z}/2\pi$ = 186.02$\pm$0.01 kHz, $\omega_{x}/2\pi$ = 680.4$\pm$0.1 kHz, and $\omega_{y}/2\pi$ = 1020.3$\pm$0.5 kHz. The $^{40}$Ca$^{+}$ ions, when confined stably in the potential, stay with 800 $\mu$m above the surface of the SET.

The phonon laser is produced by the oscillation amplification of the single $^{40}$Ca$^{+}$ ion stimulated by two 397 nm laser beams, one of which is a red-detuned beam with detuning $\Delta_{1}/2\pi$ = -75 MHz and the other of which is a blue-detuned beam with $\Delta_{2}/2\pi$ = 30 MHz. Both of the laser beams are elaborately tuned to be with the appropriate intensity ratio $r$ = $I_{b}/I_{r}$ = 0.5 ($I_{r}$ = 370 W/m$^{2}$ and $I_{b}$ = 175 W/m$^{2}$) in order to keep the oscillation amplification stable \cite{Vahala2009}, which yields the phonon laser with the oscillation amplitude of 17.839 $\mu$m. Besides, we have a saturated 866 nm laser for repumping, as sketched in Fig. \ref{fig1}a. Since the decay from $P_{1/2}$ to $S_{1/2}$ is much larger than that from $P_{1/2}$ to $D_{3/2}$, the three-level system could be reasonably considered as a two-level system \cite{Yan2019}.

In our system, we may produce the phonon laser solely regarding the $z-$axis motional degree of freedom of the ion due to the large frequency differences in different directions. Due to the same reason, we may lock the phonon laser's oscillation frequency to $\omega_{z}$, i.e., the trap frequency along $z$ axis, by applying an appropriate injection-locking signal to AE, which has no influence on other two directions. Generally speaking, when the injection frequency $\omega_{i}$ is tuned to the locking range, the phonon laser's oscillation frequency would be fixed at $\omega_{i}$ with a very sharp bandwidth \cite{Knunz2010}, implying that the oscillation phase $\varphi$, rather than the oscillation amplitude $A$, is also locked. In this case, we describe the injection force simply as $F_{0}\sin(\omega_{i}t)$, where $F_{0}$ is the force amplitude to be detected.

The main operations of our experiment are controlled by the injection and/or squeezing signal processing system, as sketched in Fig. \ref{fig1}b. The time-to-amplitude converter (TAC) gets started from the transistor-transistor logic (TTL) signal of the photon pulses via PMT and ended at the synchronized TTL signal of the injection. The time resolution of the TAC $t_{r}$ is 10 ns, which fully satisfies the requirements of the present experiment. The phonon laser is observed from the charge-coupled-device (CCD), see the images in Fig. \ref{fig1}a. With the locking signal applied, we may observe the oscillation of the phonon laser amplifying in an approximately linear way, implying that the injection locking takes action. To acquire the values of the phase $\varphi$ and the amplitude $A$ of the oscillation, we locked the oscillation frequency of the phonon laser, and recorded the accumulated photon counts within each measurement time $t_{m}$, as indicated in Fig. \ref{fig1}c.

\section{Fitting method\vspace{-1em}}
For our purpose, we have developed an approach to fast acquire the value of $A$ by fitting the recorded photons. To this end,
we assume that $\varphi$ and $A$ are constants in $t_{m}$ by ignoring the short-time noise. Under the irradiation of the 397-nm laser beams, the scattering rate $\rho_{j}$ at time $t$ is given by \cite{Leibfried2003},
\begin{equation}
\rho_{j}(t)=\dfrac{{\Gamma}s_{j}/(4\pi)}{1 + s_{j} + 4\left[ \frac{\Delta_{j}-k_{j}{\omega_{i}}A\cos({\omega_{i}}t + \varphi)}{\Gamma}\right]^{2}},
\label{eq1}
\end{equation}
where $\Gamma$ is the decay rate of $P_{1/2}$, $k_{j}$ is the wave vector, $\Delta_{j}$ means the detuning and $s_{j}$ represents the saturation parameter. Due to two 397-nm laser beams with different detunings in our experiment, the total scattering rate $\rho(t)$ = $\rho_{1}(t)+\rho_{2}(t)$. To fit the experimental photon counts, we involved the noise, such as the projection noise, the laser power noise and etc, for which the Gaussian term $G(t)$ is introduced in fitting the curve of $P(t)$ by a convolution function,
\begin{equation}
P(t) = \alpha \rho(t) {\ast} G(t) + \beta,
\label{eq2}
\end{equation}
where $G(t)$ = $(1/\sqrt{2\pi}\sigma_{t})exp[-t^{2}/(2\sigma_{t}^{2})]$ with $\sigma_{t}$ the degree of time dispersion. $\alpha$ and $\beta$ represent the factors regarding the measurement time $t_{m}$, the fluorescence collection efficiency $\eta$, the background photons and the number of unit time intervals $n$ as well as the signal-to-noise ratio (SNR) relevant to the background light. More details about the fitting can be found in Appendix.

\begin{figure}
\includegraphics[width=8.4 cm]{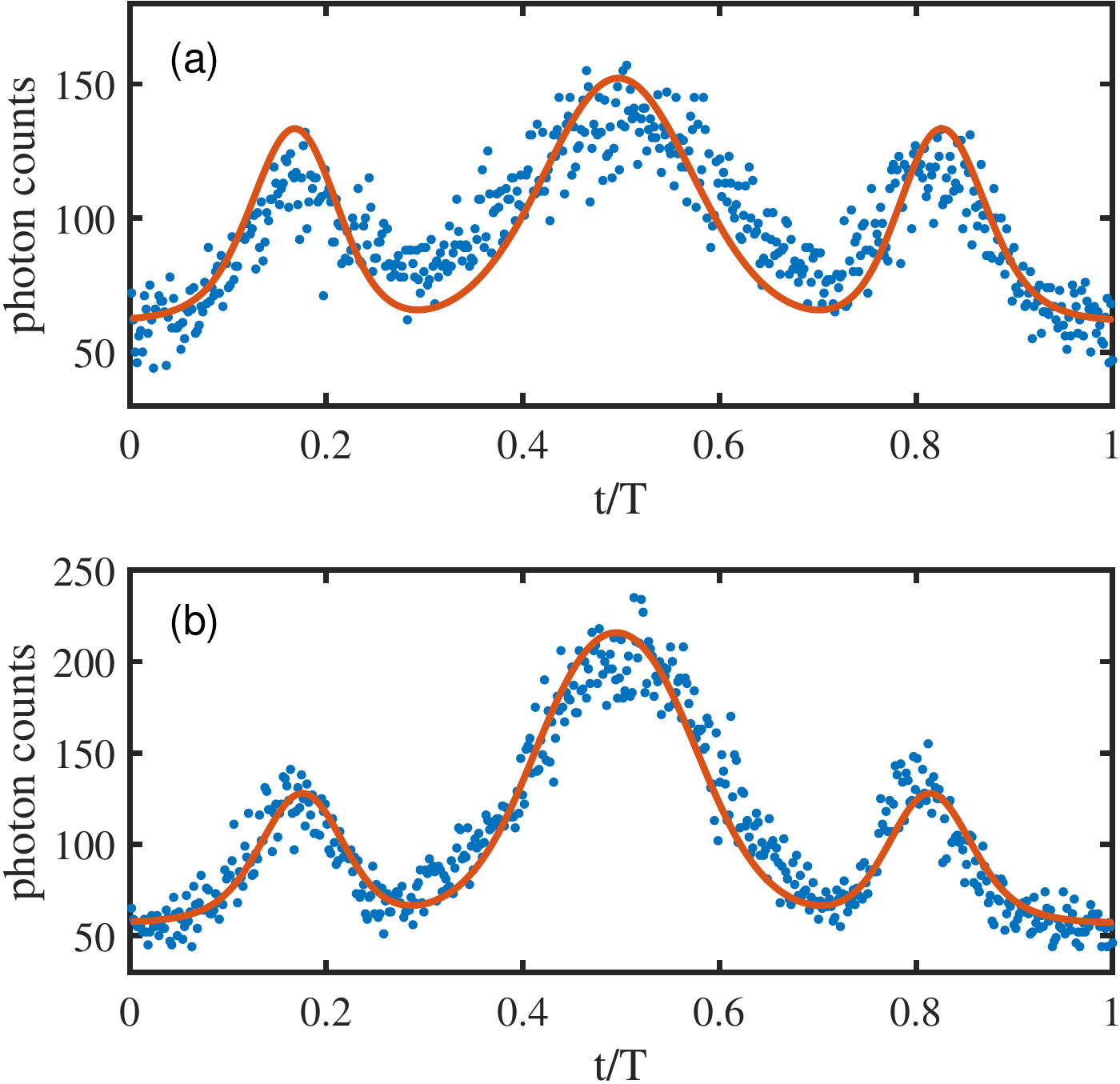}
\caption{Typical fitting cases for the accumulated photon counts. (a) $A$ = 21.677 $\mu$m and $\varphi$ = 0.028 are obtained under the injection voltage $V_{i}$ = 9.5 mV. (b) $A$ = 24.462 $\mu$m and $\varphi$ = 0.042 are acquired under the injection voltage  $V_{i}$ = 18.25 mV. The dots are experimental data and solid curves are fitted by Eq. (2). Total photon counts $N$ = 5.353$\times$10$^{4}$ are recorded by 538 unit time intervals, where each measurement time is $t_{m}$ = 10 s and we set the parameter values as $\sigma_{t}$ = 80$t_{r}$ = 0.8 $\mu$s, $\alpha$ = 4$\times$10$^{-5}$ and $\beta$ = 46. }
\label{fig2}
\end{figure}

Figure \ref{fig2} presents two fitting curves with respect to the experimental data under different injection voltages $V_{i}$. From the fitting, we acquired the oscillation amplitude $A$ and phase $\varphi$ as constants. We observed that tuning $\varphi$ mainly changes the position of the curve in the horizontal axis of $t$ and varying $A$ mainly influences the height of the second peak of the curve. This indicates that $A$ and $\varphi$ are mutually independent and thus can be acquired simultaneously by the fitting.

\section{Results and discussion\vspace{-1em}}
\subsection{Force sensing\vspace{-1em}}
The force calibration is needed prior to the force sensing due to the fact that the injection force $F_{0}$ sensed by the ion originates from a radio-frequency voltage $V_{i}$ on AE applied by the injection-locking signal and $F_{0}$ should be evaluated by comparing with the counterpart regarding a DC voltage on AE. For the latter, we have the static force $F_{c}$ = $m\omega_{z}^{2}z$ with $z$ the position deviation of the ion due to the applied DC voltage. In our experiment, we have measured $z$ = 12 $\mu$m when the applied DC voltage is 3 V, implying the force $F_{c}$ = 1088.5 zN. Thus we obtained the slope ${\partial}F_{c}/{\partial}V_{i}$ = 362.8 yN/mV.

For a harmonic oscillator, such as the trapped ion, however, the oscillation amplitude variation due to the applied radio-frequency voltage $V_{i}$ of the injection signal originates from the resonance absorption of the energy. But we may reasonably consider the amplitude variation to be driven by the force $F_{0}$ by assuming $F_{0}=F_{c}$. Since the oscillation amplitude $A$ can be accurately measured, as discussed above, by fitting the fluorescence curve when the oscillation frequency is locked, we obtained a linear relationship between the injection voltage $V_{i}$ and the oscillation amplitude $A$, with the slope ${\partial}A/{\partial}V_{i}$ = 362.1 nm/mV, see Fig. \ref{fig3}a. Excluding the original oscillation amplitude without injection, i.e., 17.839 $\mu$m, we found the relationship between the injection force and the oscillation amplitude as ${\partial}A/{\partial}F_{0}$ = 0.9979 nm/yN.
Due to frequency drifts, the measured secular frequency along $z$ axis is $\omega_{z}/2\pi$ = 186.02$\pm$0.01 kHz, which yields the uncertainty of the measured oscillation amplitude to be 15$\pm$2 nm regarding $V_{i}$. Therefore, the sensitivity of our force sensor is 347$\pm$50 yN/$\sqrt{Hz}$.

\subsection{Squeezing\vspace{-1em}}
Squeezing is an effective way to reduce the noises in precision measurement \cite{Natarajan1995,Burd2019,Majorana1997,Briant2003,Wollman2015}. In our experiment, we try to further improve the measurement precision by employing the approach of squeezing to suppress thermal noises regarding the oscillator, as explored in \cite{Natarajan1995}. Compared to the quantum mechanical squeezing to suppress quantum noises for beating standard quantum limit, the squeezing employed here is the classical squeezing, which reduces the effect of thermal noise by redistributing the thermal fluctuations. Since this squeezing is produced also by doubling the motional frequency, similar to the quantum counterpart, we may simply call it squeezing here. In our case with the Doppler-cooled ion, this squeezing helps us improve the precision of detection in the measurement of the smallest force.

The squeezing signal is applied on the EEs with frequency doubling and synchronizing with respect to the injection signal. So only the motion along $z$ axis could feel the squeezing effect. Mathematically, the uncertainty of the oscillation can be expressed by the measured standard deviation of $A$ and $\varphi$, i.e., $\sigma_{X}$ = $\sigma_{A}$ and $\sigma_{Y}$ = $A_{0}\sigma_{\varphi}$, where $A_{0}$ is the oscillation amplitude in the absence of noises. By ignoring the second-order terms in derivative as well as the higher-order terms of frequency, we acquired the variance of $Y(t)$ as \cite{Majorana1997,Briant2003},
\begin{equation}
\sigma_{Y}^{2}(g,\phi) = \frac{\sigma_{Y}^{2}(0)}{1 - g\cos(2\phi)},
\label{eq3}
\end{equation}
where $\sigma_{Y}^{2}(0)$ means the variance without squeezing and $g$ is relevant to the trapping potential modification regarding the squeezing signal (see Appendix for details).

\begin{figure}
\includegraphics[width=8.8 cm]{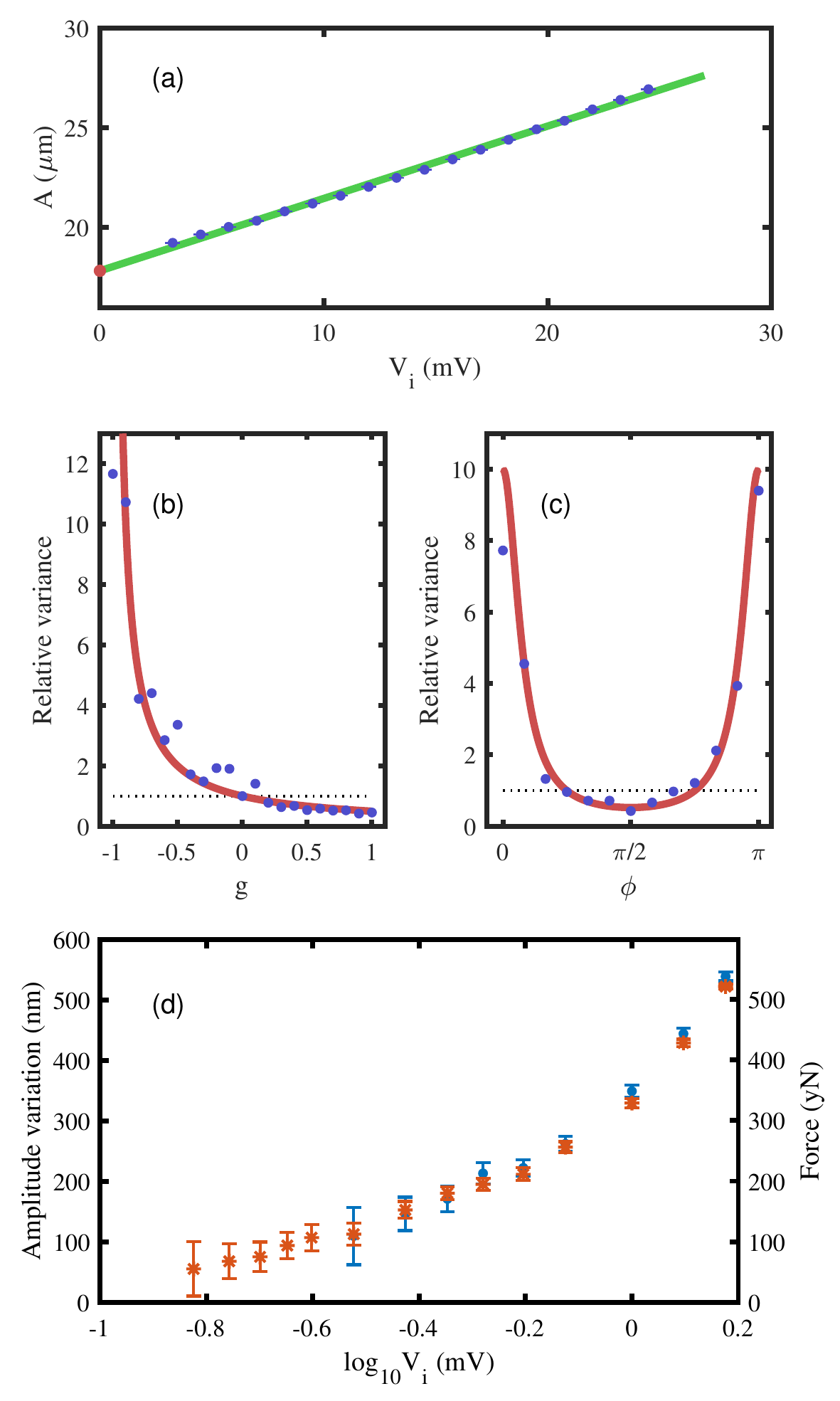}
\caption{Detection sensitivity and smallest force measurement. (a) Oscillation amplitude as a function of the injection voltage, where the red dot represent the case without the injection voltage and the blue dots are experimental data. (b) Relative variance as a function of $g$ in the case of $\phi$ = $\pi/2$. (c) Relative variance as a function of $\phi$ in the case of $g$ = 0.9. In (b,c) the variance without squeezing is normalized as unit, as represented by the dashed lines. (d) Experimental results of the lower bound of the force detection, where the blue dots are experimental data without squeezing and the red dots with 3 dB squeezing. The horizontal axis is with log$_{10}$ scale for distinguishing the difference due to the applied squeezing. The values of force and amplitude variation are identified from the measured injection voltage, following the linear relationship in the text. In all the panels, the voltage regarding the injection signal is 18.25 mV. Further voltage of 125 mV is applied on the EEs to generate 3 dB squeezing. Each measurement takes 10 seconds and we make 50 measurements for each data point. }
\label{fig3}
\end{figure}

We focus on the variance of the oscillation phase below since we have only locked the oscillation phase, rather than the oscillation amplitude. Squeezing the variance of the phase requires reducing $g\cos(2\phi)$, as indicated in Eq. (\ref{eq3}), which is achieved by tuning $g$ and $\phi$, respectively. Fig. \ref{fig3}(b,c) present the relative variance $\sigma_{Y}^{2}(g,\phi)/\sigma_{Y}^{2}(0)$ as functions of $g$ and $\phi$, where the red solid curves represent the theoretical simulation using Eq. (\ref{eq3}) and the blue dots are the experimental results with each dot acquired by 50 measurements. From the results, we identified the best squeezing effects occurred at 3 dB when $g\cos(2\phi)$ = -1.

In our experiment, we have observed the effect of squeezing on the oscillation phase, rather than the oscillation amplitude. This is due to the fact that the squeezing is relevant to the injection-locking we applied, which fixed the phase, but not the amplitude. Compared to the frequency/phase locking, locking amplitude is more challenging in ion-trap systems. For example, the feedback-locking is in principle available to fix the oscillation amplitude, but hard to achieve using current techniques since it requires higher scattering rate and higher fluorescence collection efficiency. Nevertheless, with only frequency/phase locked, we are able to reach the lower bound in the force detection, as elucidated below.

\subsection{Lower bound of force detection\vspace{-1em}}
To detect the smallest force, we may reduce the injection voltage $V_{i}$ until the phonon laser is unlocked, which is judged by fitting the values regarding the variation of the phonon laser's oscillation amplitude with respect to the voltage. Qualitatively, we identify the phonon laser working in a critical point when the oscillation frequency is locked with success probability of 90$\%$. This implies the failure of the force detection once the locking probability is lower than 90$\%$. By this way, we have detected in the absence of squeezing the critical voltage of 0.3 mV, corresponding to the lower bound of the detectable force as 171.7 $\pm$ 74.3 yN, where the large uncertainty comes from the very weak locking at the critical point and can be resorted to the error bars, as marked in Fig. \ref{fig3}(d). While with 3 dB squeezing applied, the lower bound goes further down to 86.5 $\pm$ 70.1 yN due to the successful reach of the critical voltage of 0.15 mV, where the uncertainty is also reduced due to squeezing. Since 3 dB is the best squeezing value in our experiment, we consider this lower bound to be the smallest force we can detect.

Note that, the smallest force detection in our method is determined by the frequency locking and squeezing. Repetition of measurement just helps reducing the detection uncertainty, but irrelevant to the lower bound. In this context, we have measured the electric noise on the electrodes for further understanding of the force detection lower bound. We found that the electric noise is approximately 2 mV. This fact explains the lower bound we have reached in our force detection, that is, the injection voltage could be sensed by the phonon laser only when larger than 15$\%$ (8$\%$) of the total electric noise in the absence (presence) of squeezing. This indicates that, besides reducing the measurement variance, the squeezing also improves the signal detection from the electric noise, favoring the sensing of the smaller force. In this context, to further detect a smaller force, we will have to optimize our method by suppressing the electric noises. Besides, reducing the frequency drift would be another important effort in the future. We have found that the frequency $\omega_{z}$ drifts for about 10 Hz during each measurement of 500 s. Since the locking range is really narrow, the frequency drift has the strong influence on the smallest force detection which occurs at the critical point of the frequency locking.

\section{Conclusion\vspace{-1em}}
In summary, we have demonstrated experimentally an ultrasensitive force sensor of the phonon laser of a trapped ion, where injection-locking favors high sensitivity of force detection and squeezing helps for detecting the smallest force. Our implementation worked without the prerequisite of sideband cooling, but efficiently suppressed thermal noises in the detection of the smallest force. Besides, we developed a fast approach to measure the oscillation amplitude based on counting the emitted photons. By this way, without the help of the phase Fresnel lens as employed in \cite{Blums2018}, our force detection by means of detecting the oscillation amplitude variation of the phonon laser is more sensitive to the external disturbance than by detecting the ion's position deviation. In this context, we anticipate that our scheme could be further optimized to achieve the force detection with better sensitivity in the case of higher fluorescence collection efficiency, lighter ion probe and lower electric noise. Further efforts in the future would be to suppress the secular frequency drift resulted from the electrodes' voltage drift and lock the oscillation amplitude of the phonon laser. Our employed technique can also be generalized to precisely detecting the force gradient and the surface noise due to electric fluctuation.

\section*{ACKNOWLEDGMENTS\vspace{-1em}}
This work was supported by Key Research $\&$ Development Project of Guangdong Province under Grant No. 2020B0303300001, by National Key Research $\&$ Development Program of China under grant No. 2017YFA0304503, and by National Natural Science Foundation of China under Grant Nos. 12074346, 12074390, 11835011, 11804375, 11804308, 91421111, 11734018.

Zhichao Liu and Yaqi Wei contributed equally to this work.

\begin{appendix}
\setcounter{equation}{0}
\renewcommand{\theequation}{A\arabic{equation}}
\section*{Appendix A: Dynamics\vspace{-1em}}
For a harmonic oscillator with the oscillation frequency $\omega$ under perturbation of the thermal noise, the motion is described as $z(t)$ = $[A_{0} + \delta_{A}(t)]\sin [\omega{t} + \delta_{\varphi}(t) ]$, which can be rewritten as \cite{Natarajan1995},
\begin{equation}
z(t) = X(t)\sin(\omega{t}) + Y(t)\cos(\omega{t}),
\label{eqs1}
\end{equation}
with two orthogonal components $X(t)$ = $A_{0}+\delta_{A}(t)$ and $Y(t)$ =  $A_{0}\delta_{\varphi}(t)$.
In the varying electric field, the trap secular frequency $\omega_{z}$ in our experiment is given by
\begin{equation}
\omega_{z}^{2} = \frac{Q}{m}\dfrac{\partial^{2}\psi(z)}{\partial{z}^{2}},
\label{eqs2}
\end{equation}
where $Q/m$ is the charge-to-mass ratio of $^{40}$Ca$^{+}$ ion and $\psi(z)$ is the potential along $z$ axis without squeezing.

When the classical squeezing is on, the squeezing with frequency $2\omega_{i}$ would modify the original potential, meaning $\psi(z)[1+g_{0}\sin(2\omega_{i}t+2\phi)]$, where $g_{0}$ is the assumed gain and $\phi$ is the relative phase with respect to the injection signal. In this case, the trap frequency is rewritten as $\omega_{s}^{2}(t)$ = $\omega_{z}^{2}[1+g_{0}\sin(2\omega_{i}t+2\phi)]$. For convenience of calculation, we assume $g_{0}$ = $g\zeta/\omega_{z}$, where $g$ is dimensionless and $ g\geqslant{0}$, and $\zeta$ is the frictional damping. Considering two 397 nm laser beams, we have the total frictional damping $\zeta$ = $\zeta_{1} + \zeta_{2}$ with the mathematical expression as \cite{Leibfried2003},
\begin{equation}
\zeta_{j} = -\dfrac{4{\hslash}k_{j}^{2}\Delta_{j}/\Gamma}{[1+s_{j}+4(\Delta_{j}/\Gamma)^{2}]^{2}},
\label{eqs3}
\end{equation}
where $k_{j}$, $\Delta_{j}$ and $s_{j}$ are, respectively, the wave number, detuning and saturation parameter of the $j$th beam of the 397-nm laser.

As the oscillation frequency of the phonon laser is locked and the squeezing is on, we have $\omega=\omega_{i}$ and then the Langevin equation is given by \cite{Rugar1991},
\begin{equation}
\ddot{z}(t) + \zeta\dot{z}(t) + [\omega_{z}^{2} + g\zeta\omega_{z}\sin(2\omega_{i}t + 2\phi)]z(t) = \dfrac{F(t)+f_{n}(t)}{m},
\label{eqs4}
\end{equation}
where the random force by thermal noise is $f_{n}(t) =  f_{y}(t)\sin(\omega_{i}{t})+f_{x}(t)\cos(\omega_{i}{t})$ and the injection force is $F(t)$ = $F_{0}\sin(\omega_{i}{t})$. When the oscillation frequency is locked, we have $\omega$ = $\omega_{i}$. Rewriting Eq. (\ref{eqs4}) by two orthogonal components $X(t)\sin(\omega_{i}{t})$ and $Y(t)\cos(\omega_{i}{t})$, we obtain,
\begin{equation}
\dot{X}(t)+\frac{\zeta}{2}\left(1+g\cos(2\phi)\right)X(t) = \frac{f_{x}(t)}{2\omega_{z}},
\label{eqs5}
\end{equation}
\begin{equation}
\dot{Y}(t)+\frac{\zeta}{2}\left(1-g\cos(2\phi)\right)Y(t) = \frac{F_{0} + f_{y}(t)}{2\omega_{z}},
\label{eqs6}
\end{equation}
where the second-order terms and '$\omega_{i}{t}$' terms are neglected. Thus the variances of $X(t)$ and $Y(t)$ related to temperature are $\sigma_{X}^{2}(0)$ =  $\sigma_{Y}^{2}(0)$ = $\frac{k_{B}T}{2m\omega_{i}^{2}}$ \cite{Majorana1997}. In our experiment, we just focus on $Y(t)$, whose variance with the squeezing gain $g$ can be written as in Eq. \ref{eq3} \cite{Briant2003}.

From Eq. (\ref{eqs1}), we have $\sigma_{Y}$ = $A_{0}{\sigma}_{\varphi}$, where $\sigma_{\varphi}$ is the uncertainty of the oscillation phase. Therefore, we obtain $\sigma_{\varphi}^{2}(g,\phi)/\sigma_{\varphi}^{2}(0)$ = $1/[1-g\cos(2\phi)]$. In the case of 0$< g\cos(2\phi)<$1, the uncertainty increases, whereas for -1$<g\cos(2\phi)<$0, the uncertainty reduces. We have observed experimentally that the best squeezing is 3 dB, which occurs in $g~{\rightarrow}$ 1 and $\phi$ = $\pi/2$.

\section*{Appendix B: Fluorescence collection efficiency\vspace{-1em}}
Improving the fluorescence collection efficiency of the force sensor is an important technique to achieve higher sensitivity. In our experiment, the numerical aperture of the imaging lens set is 0.36 and the transmittance is 80$\%$. The transfer efficiency of the photons by PMT is 23$\%$ and only 70$\%$ light goes through the beam splitter mirror. There are other factors in our system for the light loss of 12$\%$. For the phonon laser, the TAC transfer efficiency of the PMT pulses to TTL is about 66$\%$. Therefore, the total fluorescence collection efficiency $\eta$ is about 0.25$\%$ in theoretical prediction, which basically agrees with the measurement result 0.28$\%$. The SNR is about 2.0 and the background noise is mainly from the diffuse reflection of the two 397-nm laser beams.

The efficiency is an important factor for sensitivity. With high efficiency, the total measurement time $\tau$ could be shorten, where $\tau$ can be written as the product of the single measurement time $t_{m}$ and the measurement repetition $\varepsilon$, namely $\tau$ = $\varepsilon{t_{m}}$. In our experiment, $\varepsilon$ = 50 and $t_{m}$ = 10 s. Therefore, the sensitivity is given by \cite{Campbell2017},
\begin{equation}
S=\frac{\delta{A}\sqrt{\tau}}{\partial{A}/\partial{F_{0}}},
\label{eqs8}
\end{equation}
where $\delta{A}$ is the standard deviation of the measured oscillation amplitude. In our experiment, we have $\partial{A}/\partial{F_{0}}$ = 0.9979 nm/yN and $\delta{A}$ = 15 nm, so that the sensitivity is 347 yN/$\sqrt{Hz}$, which could be improved by lowering the total measurement time $\tau$, i.e., enhancing the fluorescence collection efficiency.

\section*{Appendix C: Drift of the secular frequency\vspace{-1em}}
When the oscillation frequency $\omega$ of the phonon laser is locked to the injection frequency $\omega_{i}$, i.e., $\omega$ = $\omega_{i}$, the bandwidth of the oscillation frequency turns to be very sharp. The locking range $\omega_{m}$ is approximately in linear proportion to the injection force $F_{0}$ and the central frequency of the locking range is exactly the secular frequency $\omega_{z}$ \cite{Knunz2010}. However, there exist drifts of the voltages applied on the electrodes, which lead to the secular frequency drifts. We have observed experimentally the drift of the frequency for about 10 Hz per 500 s. This drift has strong influence on the smallest force detection which occurs at the critical point of the frequency locking. Further suppression of such drifts would make a smaller force detection available.

\section*{Appendix D: Fitting parameters\vspace{-1em}}
Before acquiring the oscillation amplitude $A$ and phase $\varphi$ by fitting experimental data, we need to determine other parameters. The detunings of the 397-nm laser beams can be adjusted by acoustic optical modulators, and the resonant frequency is obtained by the Lorentzian curve \cite{Li2020,Urabe1998}.
The densities of the 397-nm laser beams are directly measured by saturation spectra, which are $s_{1}$ = 0.8 and $s_{2}$ = 0.4 in our experiment. The wave vectors are approximately considered as $k_{1}~\approx~k_{2}~\approx~k$, related to the resonant frequency 755.22 THz. The decay rate is $\Gamma/2\pi$ = 20.68 MHz.

Other three parameters $\alpha$, $\beta$ and $\sigma_{t}$ are obtained by the fitting. We set $\sigma_{t}$ = 0.8 $\mu$s in terms of the system's noise, and observed the oscillation amplitude 22 $\mu$m of the phonon laser. The emitted photons are theoretically 1.271$\times$10$^{7}$ within 10 seconds, while the experimentally collected photon count $N$ is 5.353$\times$10$^{4}$. Since the SNR is about 2.0, we have $\eta$ to be 0.28$\%$. Then we acquired the values of $\alpha$ and $\beta$ by $\alpha$ = $\eta{t_{m}}/n$ and $\beta$ = $N/[n(1+SNR)]$, which led to the values of $A$ and $\varphi$.

\end{appendix}

%\end{spacing}

\end{document}